# Ionic Tuning of Cobaltites at the Nanoscale

Dustin A. Gilbert,[1,2*] Alexander J. Grutter,[1*] Peyton D. Murray,[3] Rajesh V. Chopdekar,[4,5] Alexander M. Kane,[4] Aleksey L. Ionin,[4] Michael S. Lee,[4] Steven R. Spurgeon,[6] Brian J. Kirby,[1] Brian B. Maranville,[1] Alpha T. N'Diaye,[5] Apurva Mehta,[7] Elke Arenholz,[5] Kai Liu,[3, 8] Yayoi Takamura,[4] and Julie A. Borchers[1]

[1] *NIST Center for Neutron Research, National Institute of Standards and Technology, Gaithersburg, MD 20899*

[2] *Department of Materials Science and Engineering, University of Tennessee, Knoxville, TN 37996*

[3] *Physics Department, University of California, Davis, CA 95616*

[4] *Department of Materials Science and Engineering, University of California, Davis, CA 95616*

[5] *Advanced Light Souce, Lawrence Berkeley National Laboratory, Berkeley, CA 94720*

[6] *Physical and Computational Sciences Directorate, Pacific Northwest National Laboratory, Richland, WA, 99352*

[7] *Stanford Synchrotron Radiation Lightsource, SLAC National Accelerator Laboratory, Menlo Park, CA 94025*

[8] *Physics Department, Georgetown University, Washington, DC 20057, USA*

*dagilbert@utk.edu, *alexander.grutter@nist.gov

**Abstract:** Control of materials through custom design of ionic distributions represents a powerful new approach to develop future technologies ranging from spintronic logic and memory devices to energy storage. Perovskites have shown particular promise for ionic devices due to their high ion mobility and sensitivity to chemical stoichiometry. In this work, we demonstrate a solid-state approach to control of ionic distributions in $(La,Sr)CoO_3$ thin films. Depositing a Gd capping layer on the perovskite film, oxygen is controllably extracted from the structure, up-to 0.5 O/u.c. throughout the entire 35 nm thickness. Commensurate with the oxygen extraction, the Co valence state and saturation magnetization show a smooth continuous variation. In contrast, magnetoresistance measurements show no-change in the magnetic anisotropy and a rapid increase in the resistivity over the same range of oxygen stoichiometry. These results suggest significant phase separation, with metallic ferromagnetic regions and oxygen-deficient, insulating, non-ferromagnetic regions, forming percolated networks. Indeed,

X-ray diffraction identifies oxygen-vacancy ordering, including transformation to a brownmillerite crystal structure. The unexpected transformation to the brownmillerite phase at ambient temperature is further confirmed by high-resolution scanning transmission electron microscopy which shows significant structural - and correspondingly chemical - phase separation. This work demonstrates room-temperature ionic control of magnetism, electrical resistivity, and crystalline structure in a 36 nm thick film, presenting new opportunities for ionic devices that leverage multiple material functionalities.

## I. INTRODUCTION

Tailoring oxygen distributions within ionic crystals has emerged as a promising approach for controlling functional material properties, enabling new computing and energy technologies including memristors[1,2], magneto-ionic switches[3-10], and fuel cells[11,12]. While most ionic devices control electronic and magnetic properties, opportunities to control additional functionalities exist, including optical, thermal and mechanical properties. Perovskite ($ABO_3$) and perovskite-like structures are particularly attractive for ionic technologies due to their high oxygen conductivity,[11,13] and wide range of functional properties. Tuning these materials is frequently achieved through strain or interface engineering,[14-16] tailoring bond distances and angles,[17-19] or cation doping[20]. Since all these parameters are sensitive to the local oxygen environment, control may be realized through careful design of oxygen ion distributions.[21-24] Furthermore, the charged oxygen ions are ideal for mediating non-volatile electric field control of material properties.[3-5] Recently we demonstrated quasi-bulk[24] and interfacial[6] control over oxygen ion distributions, with commensurate control over magnetic ordering, using an interfacial redox technique. Other works have demonstrated local control of ion distributions in perovskites with high-energy electron beams[12,25-27] or ionic liquid techniques.[10] This solid-state approach is fast, highly scalable, and non-destructive, resolving the challenges of scalability and sample degradation[7,28] which persist with these other techniques, making it attractive for device applications. Cobalt-based perovskites possess particularly high oxygen mobility[8,11,13,29-31], making them controllable by a similar approach. Further, their tunable magnetic and electronic properties[8,32-34] make them desirable candidates for ionic devices.

In this work, oxygen ion distributions in $La_{0.67}Sr_{0.33}CoO_3$ (LSCO) films are controlled by a room-temperature solid-state redox reaction induced by a Gd capping layer. By varying the Gd thickness, we incrementally remove up to 17% of the oxygen throughout the 36 nm thick film, continuously reducing the Co valence and magnetization. In contrast, resistivity measurements show a discrete increase in the electrical resistivity. Observation of a continuous magnetic transition in the bulk measurements alongside a discrete electronic transition suggests the presence of magneto-electric phase separation and percolated conduction networks. Chemical/structural phase separation, resulting from oxygen vacancy ordering, offers an explanation for the observed magnetic and electronic results. X-ray diffraction (XRD) and high-angle annular dark field scanning transmission electron microscopy (STEM-HAADF) images confirm the presence of a structural phase separation and identify an emergent brownmillerite phase. The combined use of local measurements, which show a highly intermixed, inhomogeneous phase distribution, with quasi-bulk measurement techniques yields a unique understanding of the ion migration mechanics, and its impact on material properties.

## II. EXPERIMENT

Thin films of LSCO (≈36 nm) were grown on [001]-oriented $(La_{0.18}Sr_{0.82})(Al_{0.59}Ta_{0.41})O_3$ (LSAT) substrates by pulsed laser deposition at 700 °C in 40 Pa oxygen using a KrF excimer laser (248 nm wavelength, 1 J/cm$^2$, 1 Hz repetition),[35] then cooled to room-temperature in 40 kPa oxygen. The lattice mismatch between the LSAT (3.873 Å) and LSCO (bulk 3.834 Å), a 2.7% difference, provides a tensile strain on the LSCO. The thermal expansion of the substrate between room temperature and the deposition temperature is calculated to be 0.34%, and thus is not expected to contribute significantly to the observed effects. Polycrystalline films of Gd(0.5 nm, 1 nm, 3 nm, 5 nm)/Au(3 nm) were deposited on the LSCO by sputtering in a 0.7 Pa Ar. The Au cap was included to ensure that oxygen did not enter the film from external sources. There was no active heating of the substrate before, during or after deposition and the imparted energy from during sputtering (rate of ≈3 Å/s) is expected to constitute a negligible thermal increase. X-ray diffraction measurements were performed using a 1.5406 Å Cu $K_{\alpha 1}$ source and at the Stanford Synchrotron Radiation Lightsource, beamline 7-2, using 0.8846 Å X-rays. Plots of the XRD data are presented with coordinate axes identifying the LSAT (00$l$) Miller indices, related to $\theta/2\theta$ by $r.s.u. \equiv 3.873/d = 3.873 \times 2 \sin(\theta)/\lambda$, where $\lambda$ is the X-ray wavelength. X-ray absorption / magnetic circular dichrosim (XA/XMCD) were performed at the Advanced Light Source beamline 4.0.2. XMCD was measured using alternating circularly polarized X-rays at the Co $L_{2,3}$ edge[36] after field cooling to 110 K in a 0.5 T in-plane magnetic field. Polarized neutron reflectometry (PNR) was performed at the NIST Center for Neutron Research on the MAGIK and PBR reflectometers at 110 K, with a +30 mT in-plane magnetic field, after field cooling in -750 mT.[37] The reflectometry data for the non spin-flip channels is presented ($R^{\text{Incident Scattered}}$: $R^{++}$, $R^{--}$). Model fitting was performed using the Refl1D software package;[38] models for the three samples were fitted in parallel with the nuclear scattering densities, $\rho_N$, of the LSAT and Au are coupled between the models. The LSCO layer in the PNR model is divided into three sublayers to allow for depth-dependent variation. Resistivity and transverse magnetoresistance (e.g. magnetic field and current in-plane and mutually orthogonal) measurements were performed at 77 K using a 4-probe setup. Samples for STEM-HAADF imaging were prepared using a lift-out method with a focused ion beam microscope operating at 4°-7° incidence angle at energies of 2 keV – 30 keV; imaging was performed at 200 keV along the LSAT [110] zone-axis with a 1 Å probe size. Images taken at 80 keV and 200 keV show the same features, suggesting that the results are not caused by beam damage. Electron energy loss spectroscopy (EELS) was measured by monitoring the oxygen $K$, Sr $L_{2,3}$, La $M_{4,5}$, and Co $L_{2,3}$ edges.

**III. RESULTS**

X-ray absorption spectra of the LSCO films, Figure 1(a), are in agreement with previous measurements on these materials.[36,39] With increasing $t_{Gd}$ the peak of the Co-$L_{3,2}$ edge shifts to lower photon energies, indicating a reduction of the Co valence from its initial nominal value of Co$^{3.33+}$.[39] For the thickest $t_{Gd}$ there is an emergent peak at 776 eV, consistent with an emergent Co$^{2+}$ phase. An increase in Co valence is consistent with oxygen migration from LSCO to Gd, as each oxygen ion is expected to return two electrons to the LSCO as the Gd is oxidized. The collated trends for all the data, including the XA, are presented in the discussion section. At the peak of the $L_3$ absorption edge (779 eV) the X-ray transmission through our film is calculated to be 88%, indicating that the fluorescence yield (FY) measurements are probing the entire film thickness. This suggests that that oxygen is extracted from throughout the entire thickness of the film, which is confirmed later with neutron reflectometry and electron microscopy. The Co XMCD intensity measured at 110 K, Figure 1b, decreases continuously with increasing $t_{Gd}$, disappearing completely for the LSCO/Gd(5 nm) sample, demonstrating full suppression of the ferromagnetism in the sample. Complementary measurements taken using the surface sensitive total electron yield (TEY) signal are shown in the Supplementary Material.[40]

Polarized neutron reflectometry results provide depth resolved nuclear and magnetic profiles of the samples.[41,42] Measured reflectometry data, Figure 1(c-e), shows a strong splitting of the R$^{++}$ and R$^{--}$ channels for the Gd(0 nm) and Gd (1 nm) samples, but no splitting for the Gd(5 nm) sample. The reduction of the splitting between the R$^{++}$ and R$^{--}$ channels strongly suggests suppression of the ferromagnetism in the LSCO with increased Gd thickness, in agreement with the XMCD measurements. The converged models ($\chi^2$ of <2.5), shown in Figure 1(f-h), reproduce the expected structure and nuclear scattering length densities for LSAT, Au, and La$_{0.67}$Sr$_{0.33}$CoO$_3$ (for $t_{Gd}$=0 nm). With increasing $t_{Gd}$, $\rho_N^{LSCO}$ decreases throughout the entire film, with $\rho_{N,meas}^{LSCO}$ ($t_{Gd}$=0 nm)=4.87×10$^{-4}$ nm$^{-2}$, $\rho_{N,meas}^{LSCO}$ ($t_{Gd}$=1 nm)=4.76×10$^{-4}$ nm$^{-2}$, and $\rho_{N,meas}^{LSCO}$ ($t_{Gd}$=5 nm)=4.47×10$^{-4}$ nm$^{-2}$ (uncertainty of <±0.01×10$^{-4}$ nm$^{-2}$). From $\rho_{N,meas}^{LSCO}$ the oxygen stoichiometry and Co valence are calculated[5] to be La$_{0.67}$Sr$_{0.33}$CoO$_{3.00}$ (Co valence of 3.33), La$_{0.67}$Sr$_{0.33}$CoO$_{2.88}$ (Co valence of 3.09), and La$_{0.67}$Sr$_{0.33}$CoO$_{2.47}$ (Co valence of 2.27) for $t_{Gd}$ = 0 nm, 1 nm, and 5 nm, respectively (propagated error on the oxygen stoichiometry is ±0.01). At $t_{Gd}$ = 5 nm, 17% of the oxygen (0.53 O per u.c.) has been extracted from the LSCO. The PNR results also act to verify that we do not have significant Gd diffusion into the LSCO. As Gd is a strong neutron absorber, diffusion would appear in the imaginary SLD as a broadening at the interface, which we do not observe.

Commensurate with the decrease in $\rho_N^{LSCO}$ is a decrease in the magnetic SLD, $\rho_M^{LSCO}$, which is proportional to the magnetization. The $\rho_M^{LSCO}$ of the LSCO/Gd(0 nm) sample indicates a saturation magnetization ($M_S$) of 104 emu/cm$^3$ (1 emu/cm$^3$ = 10$^3$ A/m; 104 emu/cm$^3$ = 0.64 $\mu_B$/Co), consistent with previous works.[43] Depositing Gd(1 nm) reduces $\rho_M^{LSCO}$ by 20%, while $\rho_M^{LSCO}$≈0 for the LSCO/Gd(5 nm) sample, indicating no ferromagnetic (FM) ordering. Along with $\rho_N^{LSCO}$, $\rho_M^{LSCO}$ decreases throughout the entire film demonstrating high oxygen mobility in the LSCO layer.[8,11,13,29-31] The continuous decrease in the Co valence determined from $\rho_{N,meas}^{LSCO}$ is consistent with the XA results, while the suppression of the magnetization determined from $\rho_M^{LSCO}$, is consistent with the XMCD. We note that the Gd is not magnetic in any of the PNR profiles, as would be expected even for these thin Gd films at 110 K,[44] thus suggesting the Gd is fully oxidized. With the Au cap preventing oxidation from external sources, oxygen is expected to come from the LSCO underlayer.

Transverse magnetoresistance (MR) measurements taken at 77 K, Figure 1(i), show the expected shape, with the resistance peak at ≈1.25 T identifying magnetization reversal. The anisotropic MR measured in this configuration increases as $R \propto \cos(\theta)$, where $\theta$ is the angle between the magnetization and the current. Thus, the MR is expected to be minimized at saturation, and maximal at the coercive field. The MR peak position is consistent with the expected large magnetocrystalline anisotropy ($K_U$) of LSCO,[45] but interestingly does not change for $t_{Gd} \leq 1$ nm. The electrical resistivity, Figure 1(j), is also approximately constant over this range. The as-grown sample shows a resistivity of $4\times10^{-4} \pm 5\times10^{-5}$ Ω cm, consistent with expected values. Strikingly, the LSCO/Gd(5 nm) sample exhibits high resistivity and no MR. The resistivity of the LSCO is measured in parallel with the Gd/Au capping layers, and thus the measured 34× increase is an *underestimate* of the actual change in resistivity, however, the Gd/Au cap means that the detailed temperature dependent resistance measurements necessary to confirm a metal-insulator transition cannot be performed. Thus, the MR and resistivity measurements show that, as oxygen is removed from the LSCO, there is a rapidly progressing transition from the as-grown metallic, FM phase[33] to a much higher resistance, non-ferromagnetic phase.

The MR peak position tracks the coercivity, which scales with $K_U/M_S$; XMCD and PNR show $M_S$ decreases with increasing $t_{Gd}$, implying the MR peak position should increase. The surprising insensitivity of the MR peak position and resistivity to Gd thickness for $t_{Gd} \leq 1$ nm is consistent with phase separation occurring within the film, as one possible explanation.[46] A high anisotropy FM low-resistance phase coexisting alongside a non-ferromagnetic highly-resistive phase would exhibit a lower average magnetization while retaining a high coercivity. Furthermore, for highly-intermixed phases, percolation theory predicts a sharp increase in the electrical resistivity at a critical oxygen stoichiometry,[47] as seen here. While reference [46] indicate that such a transition can be driven by cation doping, electron microscopy and XRD results presented below suggest structural phase separation, manifested by oxygen vacancy ordering, is responsible in this work. Discussion of these competing explanations is provided in the discussion. The oxygen stoichiometry in the LSCO/Gd(5 nm) sample and recent work on $SrCoO_3$[11] identify the brownmillerite ($A_2B_2O_5$) structure as a likely secondary phase. Unlike perovskites, in which every $B$-site ion is surrounded by an oxygen octahedra, brownmillerites have alternating planes of octahedra and tetrahedra along the [001] direction, as illustrated in Figures 2(a) and (b), respectively. However, the brownmillerite phase possesses entirely different material properties than the perovskite structure, with $La_{1.33}Sr_{0.66}Co_2O_5$ previously reported as a non-ferromagnetic insulator,[10] consistent with our MR and resistivity measurements.

High-resolution XRD for the LSCO/Gd(0 nm) sample, Figs. 2(c, d), shows only the (00$l$) family of peaks for the LSAT and LSCO, flanked by Kiessig fringes, confirming high-quality growth of the initial perovskite-structured films. From the (00$l$) peaks, the $c$-lattice parameter of the as-grown LSCO film was calculated to be 3.797±0.002 Å, smaller than the bulk value of 3.834 Å due to the coherent tensile strain.[48] As $t_{Gd}$ increases, the LSCO peak continuously shifts to lower angles while maintaining constant width, indicating an increase in the $c$-lattice parameter. Since the XRD measurement probes the entire film thickness, this implies oxygen vacancies are likely distributed throughout the film, consistent with the PNR and XA results. XRD of the LSCO/Gd(3 nm) sample reveals two LSCO peaks at each Miller index coordinate, identifying coexistent, structurally distinct phases, with one phase being the as-grown structure, and the other likely an oxygen-deficient phase. Reciprocal space maps (RSM) [Figures 2(e-g)] confirm epitaxial growth, with the common $h$ coordinate of the film and substrate peaks indicating that the LSCO remains fully strained to the substrate regardless of $t_{Gd}$, while the shifting $l$ coordinate of the film peak indicates that only the $c$-lattice parameter changes. Additional XRD peaks and RSMs are shown in the Supplemental Materials.[49]

With increasing $t_{Gd}$, a series of superlattice peaks also appear at ½-order locations, Figure 2(c). These peaks indicate a doubling of the unit cell along the $c$-axis, consistent with the alternating octahedra/tetrahedra layering in the brownmillerite phase [Figure 2(a, b)]. While other phases possess the same chemical stoichiometry and peak locations,[50] the peak intensities suggest brownmillerite is the dominant phase; electron microscopy imaging, discussed below, confirms the brownmillerite ordering. The gradual emergence of the half-order XRD peaks indicates an evolution towards brownmillerite dominance at higher $t_{Gd}$. Previous studies of LSCO films under tensile strain have reported oxygen vacancy ordering along the in-plane direction to relieve the strain.[51-54] However, our measurements show out-of-plane vacancy ordering, suggesting that the phase transformation occurs due to a fundamentally different mechanism, with a possible mechanism presented in the discussion section.

Select area STEM-HAADF images of the LSCO/Gd(1 nm) sample, Figure 3(a), show alternating bright/dark bands, highlighting brownmillerite ordering.[26,55] The octahedra/tetrahedra layers have out-of-plane spacings of 4.07±0.05 Å and 3.77±0.05 Å, respectively, for a total unit-cell height of 7.84±0.05 Å, consistent with previous results[29,55] and the X-ray results. The unit cell doubling along the $c$-axis is identified by the half-order peaks in the fast Fourier transform (FFT). The STEM image confirms the brownmillerite ordering to be exclusively along the out-of-plane direction, consistent with the suggestion that the Gd-induced chemical potential gradient is driving the phase transformation. The depth-resolved oxygen stoichiometry of the LSCO, determined by PNR, is plotted adjacent to the STEM image, Figure 3(b), showing excellent structural agreement and confirming oxygen migration throughout the film.

The STEM and PNR measurements are complementary and give an enhanced understanding of the oxygen distribution and phase separation. Specifically, the STEM images highlight the local structural phase separation and corresponding variations in the oxygen profile, identifying a highly inhomogeneous system. EELS measurements performed in a brownmillerite region, giving a local measure of the oxygen stoichiometry, shown in Figures 3(c) and 3(d), confirm the expected profiles of O, La, Sr and Co. The Sr:O ratio is approximately 0.3±0.02 in the LSAT substrate, and 0.15±0.02 for the LSCO. Assuming the nominal stoichiometry for La, Sr and Co, the stoichiometry of the brownmillerite LSCO is calculated as $La_{0.66}Sr_{0.33}CoO_{2.3\pm0.3}$. The Sr:O plot shows a slight slope along the sample thickness, with the ratio getting larger near the top surface, suggesting a reduced oxygen stoichiometry approaching the Gd in this region of the sample. Complementary to this, the PNR depth profile shows that the average oxygen stoichiometry is relatively uniform throughout the depth, owing to the lateral phase segregation, and highlighting the high oxygen mobility. Plotting the PNR and STEM results side-by-side also helps identify a particularly oxygen deficient region at the LSCO/Gd interface. The PNR results show a gradual decrease in oxygen content, while the STEM imaging shows a significant decrease in ordering and the absence of a sharp interface. This may be due to the particularly strong reduction occurring at this interface, or the large number of oxygen ions conveyed through this region.

Close examination of the STEM-HAADF image, Figure 3(e), confirms the structural phase separation inferred by XRD and MR. Regions of perovskite and brownmillerite ordering are identified, with the intensity profiles shown in Figure 3(f). These profiles show that the brownmillerite regions possess an alternating two-amplitude periodicity, highlighting the bright/dark banding, while the perovskite region shows a single amplitude periodicity. Closer to the interface a highly deficient disordered cubic phase may exist, predicated on random spatial distributions of the oxygen vacancies.[12] The STEM images also reveal stacking faults and anti-phase boundaries which are not expected in the as-grown film, shown in Figures 3(g) and 3(h),[56] and are known oxygen migration pathways.[12] Thus the lateral separation may be the result of reduced oxygen conductivity in the brownmillerite regions.[57]

IV. DISCUSSION

The above results demonstrate the active design of oxygen ion distributions in an LSCO thin-film, realized by the deposition of a reactive Gd capping layer. As oxygen is removed electrons are returned to the Co ions, reducing its valence, causing a shift in the $L_3$ XA peak, tabulated in Figure 4(a). The removal of oxygen is also observed in the nuclear depth profile from PNR; the oxygen stoichiometry and Co valence are calculated from $\rho_N^{LSCO}$, Figure 4(b), and are in agreement with the XA. Commensurate with the removal of oxygen, the measured magnetization is suppressed, as shown in the XMCD and $\rho_M^{LSCO}$, tabulated in Figure 4(c), the $c$-lattice parameter increases, Figure 4(d), and the resistance rapidly increases, Figure 4(e).

We may gain insight into the roles of the electron doping and ion/structural effects by recalling that two electrons are returned to the Co ions for each extracted oxygen ion,[58] electron doping the LSCO. Previous works have shown that metal-to-insulator and spin-glass to FM transitions occur at a Co valence of +3.18.[34,43,59] In this work, the as-grown film is metallic and FM due to the $Sr^{+2}$ hole doping, with a Co valence of +3.33. In the LSCO/Gd(1 nm) sample, the average Co valence is reduced to +3.09, below the transition value of +3.18, implying that, if the electrons were uniformly doped throughout the film, the LSCO should be insulating and non-ferromagnetic.[32,34,43] However, metallic FM behavior persists, implying that the oxygen extraction cannot be treated as a uniform introduction of electron dopants. Thus, the structural/chemical phase separation must play a dominant role in the demonstrated material control. While we observe a suppression of the magnetization through the use of many measurements techniques, the nature of the resultant magnetic ordering was not investigated. While Reference 44 would suggest the magnetic state may include spin-glass or paramagnetic phases, a traditional brownmillerite LSCO phase is expected to show antiferromagnetic ordering. However, the sample volume is too small to investigate the antiferromagnetic ordering with high-angle neutron diffraction. Without investigating the ordering directly, we make no claims to what it may be, only noting that there is no net magnetic moment suggesting suppression of ferromagnetic ordering.

Crucial to the above control is the thermodynamics of removing the oxygen ions and inducing the brownmillerite transformation. The oxygen transport observed in this work does not require the input of significant thermal energy, as is often necessary for magneto-ionics, indicating that the LSCO/Gd reaction occurs spontaneously at room-temperature. The feasibility of a spontaneous reaction can be determined by calculating the Gibbs free energy. An extreme limit of fully-reducing LSCO to its elemental components is considered; formation of intermediate products, including the expected $La_2O_3$+SrO+Co or the introduction of isolated oxygen vacancies, is expected to be of lower-energy. Values for the enthalpy, $\Delta H$, and entropy, $\Delta S$, were estimated using the parent compound $LaCoO_3$ ($\Delta H$ = 1234 kJ mol$^{-1}$, $\Delta S$=111.3 J mol$^{-1}$ K$^{-1}$),[60] and bulk values for $Gd_2O_3$ ($\Delta H$ = 1819 kJ mol$^{-1}$, $\Delta S$=150.6 J mol$^{-1}$ K$^{-1}$).[61] The Gibbs free energy, $\Delta G = \Delta H - T\Delta S$, is calculated at room-temperature (300 K) for the extreme case of complete reduction of the LSCO, $La_{0.7}Sr_{0.3}CoO_3 + 2Gd \rightarrow La_{0.7}Sr_{0.3}Co + Gd_2O_3$, representing a high-energy limit. For the extreme case of complete reduction $\Delta G$ = -598 kJ mol$^{-1}$. The negative $\Delta G$ confirms that the Gd cap is able to fully reduce the LSCO spontaneously at room-temperature to its elemental components. The $\Delta H$ and $\Delta S$ values for $La_{0.7}Sr_{0.3}CoO_{3-\delta}$ to $La_{0.7}Sr_{0.3}CoO_3$ are expected to be less than the full reduction, further emphasizing the ability of Gd to partly reduce LSCO spontaneously at room-temperature. An implication of the energetics is that oxygen leaching causes the manifestation of a chemical potential gradient within the material. That is, as oxygen is extracted from the LSCO at the top surface (e.g. the interface with Gd), the chemical - and potentially strain - energies become inhomogeneous along the thickness of the film. The use of a continuous Gd film makes the chemical potential at any particular depth laterally uniform. The resultant uniaxial chemical potential gradient motivates the observed ion migration to occur preferentially along the thickness of the film. However, it is well known that ion migration preferentially occurs at defect boundaries,[12] which are expected to be relatively scarce in the as-grown epitaxial film. Accordingly, as oxygen is removed from the top surface there may be significant lateral ion migration within the film to these defect boundaries, thus resulting in the lateral orientation of the brownmillerite structure.

Particularly remarkable is the fact that this control is achieved throughout the entire 36 nm thickness of the film. Previously we have demonstrated redox control to design bulk[24] and interface[6] oxidation states. While bulk compared to interface design may be due to the high oxygen mobility in perovskite oxides, it has recently been suggested that the range of control is determined by the heat of formation and the electron work function.[62] In the case of LSCO/Gd, the Gd has a much lower electron work function and more negative heat of formation, allowing oxygen leaching to extend deep into the LSCO film.

## V. CONCLUSIONS

In conclusion, by harnessing the high oxygen conductivity and percolative phase transformations of perovskite systems, this work advances ionic-control of materials from limited localized control of single-material properties, to comprehensive long-range control of multiple functionalities. XA and PNR results both show that oxygen is removed throughout 36 nm thick $La_{0.67}Sr_{0.33}CoO_3$ films, reducing the Co valence and suppressing the magnetization. Magnetoresistance measurements demonstrate that as oxygen ions are removed, the coercivity remains large and the resistivity remains low until a critical oxygen vacancy threshold, implying phase separation. X-ray diffraction and STEM-HAADF imaging confirms a structural phase separation and identify a percolated brownmillerite phase coexisting alongside the perovskite phase, implying commensurate chemical phase separation. This work demonstrates control of bulk-like magnetic, structural, and electronic properties, achieved by room-temperature treatment. Coupling this work with recent advances in electric-field control of ion distributions opens new pathways towards nonvolatile control of ionic materials at the nanoscale.

**Acknowledgements**

This work was supported by the US Department of Commerce. D.A.G. and A.J.G. acknowledge support from the National Research Council Research Associateship Program. Work at UCD was supported by the NSF DMR-1610060(P.M), ECCS-1611424 (K.L.), and the University of California Multicampus Research Programs and Initiatives (MR-15-328528) (R.V.C., A.M.K., A.L.I., M.S.L. and Y.T.). S.R.S is supported by the U.S. Department of Energy (DOE), Basic Energy Sciences (BES), Division of Materials Sciences and Engineering under Award #10122. The PNNL work was performed in the Environmental Molecular Sciences Laboratory, a national scientific user facility sponsored by the Department of Energy's Office of Biological and Environmental Research and located at PNNL. This research used resources of the Advanced Light Source, which is a DOE Office of Science User Facility under contract no. DE-AC02-05CH11231. Use of the Stanford Synchrotron Radiation Lightsource, SLAC National Accelerator Laboratory, is supported by the U.S. Department of Energy, Office of Science, Office of Basic Energy Sciences under Contract No. DE-AC02-76SF00515.

**Figures**:

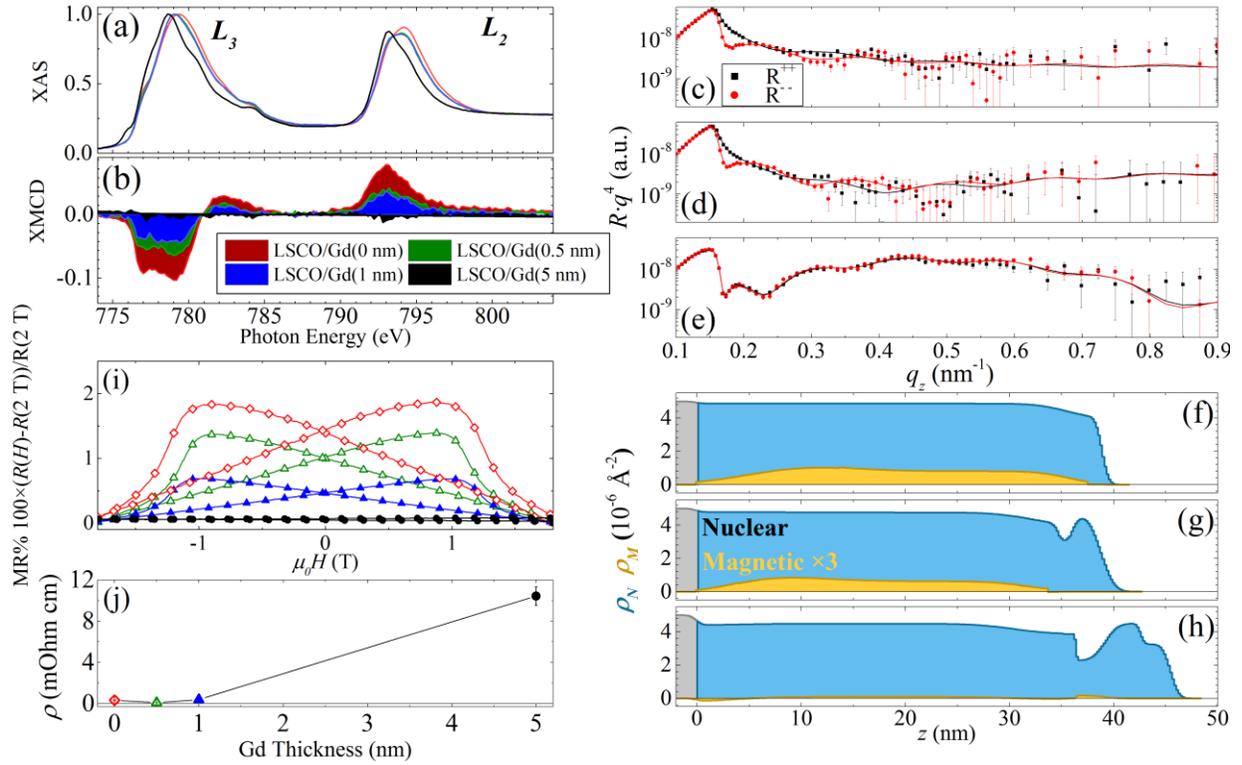

**Figure 1.** (a, b) Co-XA/XMCD spectra of LSCO/Gd($t_{Gd}$) samples. Polarized neutron reflectometry patterns (c-e) and fitted profiles (f-h) for (top to bottom) LSCO/Gd(0 nm), LSCO/Gd(1 nm), and LSCO/Gd(5 nm). (i) MR measurements of LSCO/Gd($t_{Gd}$) samples. (j) Electrical resistivity of LSCO/Gd($t_{Gd}$)/Au(3 nm) heterostructure. Error bars identify the uncertainty determined by the tool accuracy. X-ray and PNR measurements were performed at 110 K, while resistivity was taken at 77 K.

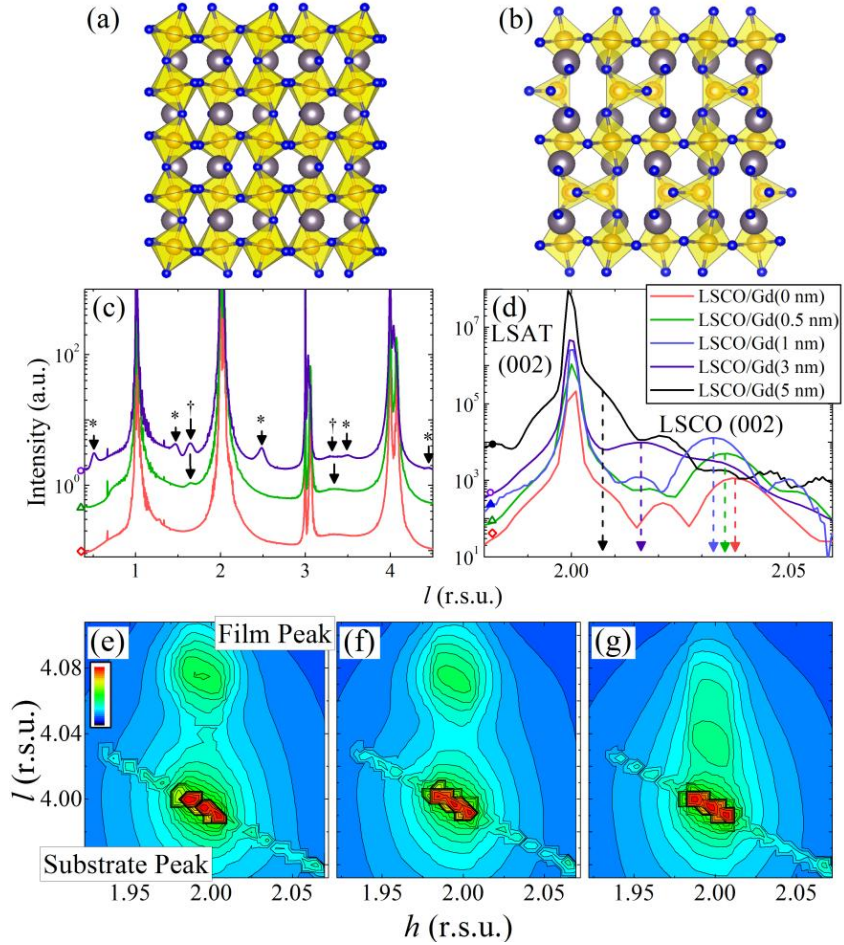

**Figure 2.** Diagrams of (a) perovskite and (b) brownmillerite crystal structures. (c) XRD patterns of the LSCO and LSAT (00$l$) family of peaks and (d) zoomed in view of the (002) peaks. Open and solid symbols indicate data taken with synchrotron and Cu $K_{\alpha 1}$ X-rays, respectively. Brownmillerite peaks are identified by (*); peaks from the Au cap are indicated by (†). Dashed lines are guides for the eye. RSMs of (e) LSCO/Gd(0 nm), (f) LSCO/Gd(0.5 nm), and (g) LSCO/Gd(3 nm) samples measured around the (204) peaks; axes are labeled with the LSAT $h$ and $l$ Miller indices.

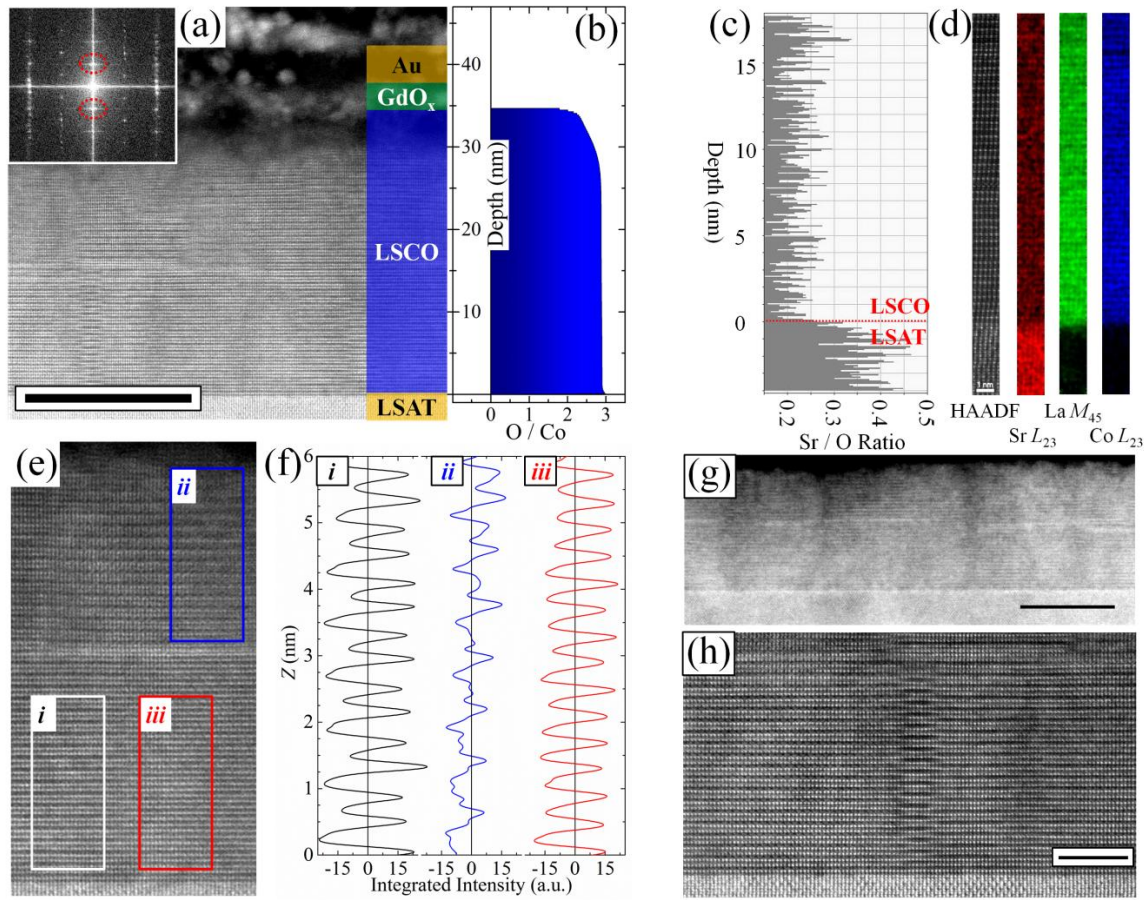

**Figure 3.** (a) Cross-sectional STEM-HAADF image of the LSCO/Gd(1 nm) sample. Inset shows the FFT of the LSCO film with the (001) oriented brownmillerite unit cell identified by the dashed red circles. (b) Oxygen stoichiometry profiles from the PNR nuclear profile. STEM-HAADF EELS map of (c) Sr:O depth profile and (d) (left to right) structural and element specific maps of the Sr, La, and Co profiles. (e) Zoomed-in image identifying coexisting (*i*) brownmillerite, (*ii*) less-ordered brownmillerite and (*iii*) perovskite phases. (f) The horizontally-integrated intensity from the boxes in (e). (g) Wide angle and (h) high-resolution STEM images of of LSCO/Gd(1 nm) highlighting the induced antiphase boundaries and stacking faults. Scale bar in (a), (g) and (h) indicate 20 nm, 20 nm and 5 nm, respectively.

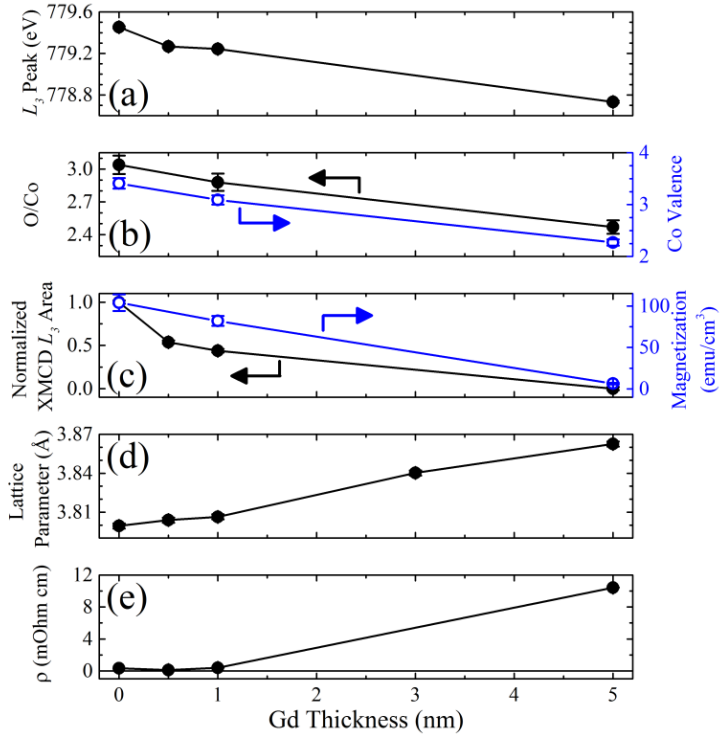

**Figure 4.** Collated trends showing the trends in (a) the L3 XA absorption peak, (b) the O/Co and Co valence state, (c) the XMCD area and magnetization from PNR, (d) the average lattice parameter and (e) the measured resistivity.